# Engineering quantum interference

M. Lucci, V. Merlo, I. Ottaviani, M. Cirillo, D. Badoni, V. Campanari, G. Salina, J. G. Caputo, and L. Loukitch

# Engineering quantum interference

M. Lucci,[1] V. Merlo,[1] I. Ottaviani,[1] M. Cirillo,[1,a)] D. Badoni,[2] V. Campanari,[2] G. Salina,[2] J. G. Caputo,[3] and L. Loukitch[3]

[1]*Dipartimento di Fisica and MINAS Lab Università di Roma "Tor Vergata," 00133 Roma, Italy*
[2]*Istituto Nazionale di Fisica Nucleare, Sezione di Roma "Tor Vergata" Via della Ricerca Scientifica 1, 00133 Roma, Italy*
[3]*Laboratoire de Mathématiques, INSA de Rouen, Avenue de l'Université, F-76801 Saint-Etienne du Rouvray, France*

A model for describing the interference and diffraction of wave functions of one-dimensional Josephson array interferometers is presented. The derived expression for critical current modulations accounts for an arbitrary number of square junctions, the variable distance between these, and the variable size of their area. Predictions are tested on real arrays containing up to 20 equally spaced and identical junctions and on arrays shaped with peculiar geometries. A very good agreement with the modulations predicted by the model and the experimental results is obtained for all the tested configurations. It is shown that specific designs of the arrays generate significant differences in their static and dynamical (non-zero voltage) properties. The results demonstrate that the magnetic field dependence of Josephson supercurrents shows how interference and diffraction of macroscopic quantum wavefunctions can be manipulated and controlled. *Published by AIP Publishing.*
https://doi.org/10.1063/1.5057767

Josephson interferometers, namely, superconductive loops embedding Josephson junctions, represent a relevant topic in fundamental and applied physics. Such systems have provided the basis for developments of quantum-limited magnetic sensors,[1] core elements[2] for the realization of fast digital circuits,[3] proposals of quantum computing concepts,[4] and superconducting meta-devices.[5] Most of these applications employ the basic properties of single interferometers in order to engineer more complex systems and devices. Feynman, Leighton, and Sands put forward the idea that parallel arrays of many Josephson interferometers could lead to potentially interesting phenomena.[6] In the last few decades, one-dimensional parallel arrays of junctions indeed have been the subject of several investigations,[7–10] but several features and properties of these systems have not been accounted for.

In addition to the interest in fundamental physics, nowadays the search for new detectors[11] and computation devices[12] has renewed the attention for Josephson interferometers and parallel arrays of Josephson junctions and we herein intend to contribute to a comprehensive understanding of multi-junctions diffractions and interference patterns which may constitute a solid background for future developments. The results on arrays of different geometries show that diffraction and interference in one-dim Josephson grids can be controlled and engineered to a high degree of accuracy.

In Fig. 1(a), we sketch a top view of the physical system being the subject of our investigations: a parallel array of square Josephson junctions. In the figure, we see that the distances between the squares, the areas of the junctions, are all different and the squares have different areas: in our model, indeed both distances between junctions and the areas of these can be arbitrary, but the areas must always refer to a square geometry. In Fig. 1(b), instead we show a section (not to scale) of the array limited to two junctions and show the areas through which the magnetic field links to the junctions and to the connecting loops: these are identified by the dashed lines enclosing the areas $dw_j$ and $Dl$. Here, $d$ represents the sum of London penetration depth of the base and top electrodes, while $D$ is essentially $d$ with the thickness of the $SiO_2$ insulating layer added to it.

Our approach to explain field modulations has been developed as a generalization of the two-junction interferometer

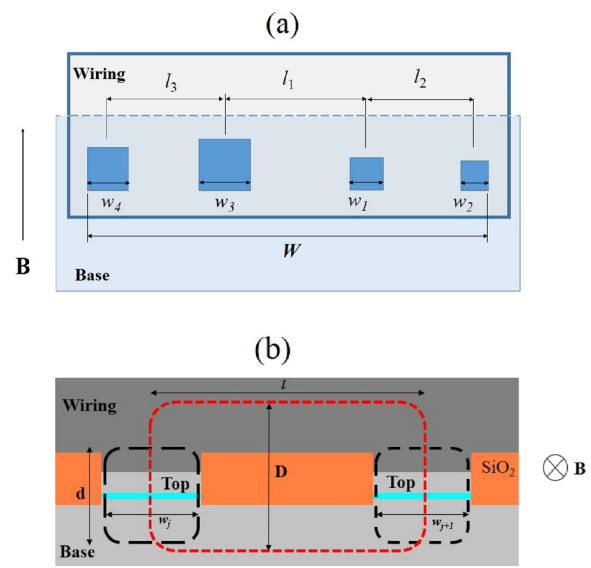

FIG. 1. (a) Sketch of the system under investigation. The squares with side $w_j$ indicate the locations of the junctions and (b) cross section of two junctions showing the areas (indicated by the dashed contours) through which the magnetic field links to junctions (smaller rectangles) and to the connecting loop (larger rectangle). Indexing of the junctions is the one we use when developing the terms of Eq. (1) for arrays with an even number of junctions.

a)Author to whom correspondence should be addressed: matteo.cirillo@roma2.infn.it

analysis,[13] previously attempted.[10] Overall, the idea is to consider a parallel array of junctions like a single junction with a step-like current dependence along one direction. An example is given in the inset of Fig. 2(a) for a six junction array: here, a junction of width $W$ is assumed to have an internal structure in which five regions of zero current (the loops) separate six regions with the given Josephson current. Herein, the geometrical shape of the junctions is always assumed as a square and therefore, from the current density (assumed constant here all over the array) and the width, the current through the junction in zero external field is given. We will calculate the effect of the external magnetic field on this distribution assuming that every individual junction has physical dimensions much smaller than the Josephson penetration depth. The Josephson current passing through the array is calculated from the spatial Fourier transform of a $J(x)$ dependence like the one in the inset of Fig. 2(a). This type of approach originated from the arguments contained in Sec. 4.4 of Ref. 1: in Ref. 13, the authors used the techniques therein described for fitting magnetic field modulations of the current of interferometers. An extension of the work presented in Ref. 13 was attempted, based on the linearity of the Fourier transform, in Ref. 10 but, only for arrays with an even number of junctions $N$; we extend the linearity principle to calculate the total currents for $N$ even or odd, and arbitrary spacing between the junctions considered of arbitrary square area.

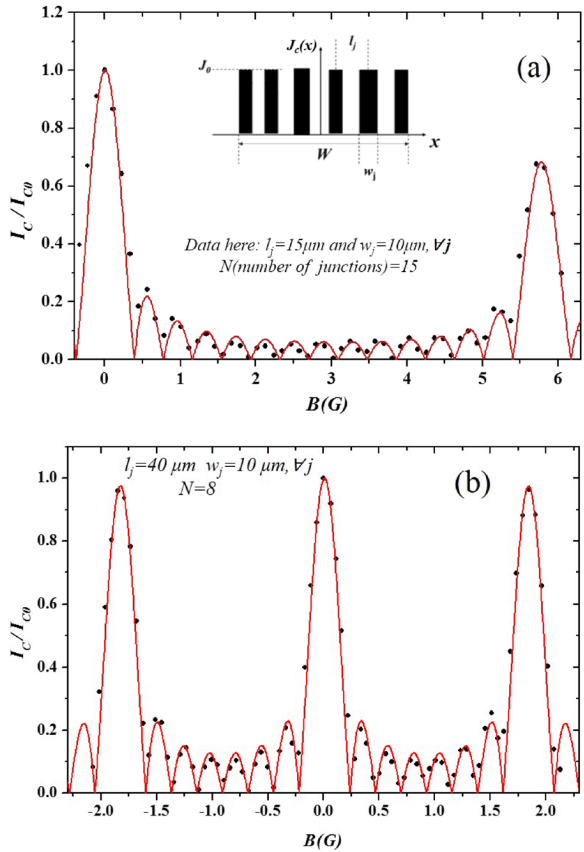

FIG. 2. Modulations of the maximum Josephson current of an array containing, respectively, 15 junctions spaced 5 $\mu$m (a) and 8 junctions spaced 30 $\mu$m (b). The curves fitting the data are the results of Eq. (1) considering these geometrical parameters. The inset in (a) indicates the concept beyond our model: we look at arrays (of six junctions in this example) as a single junction with a step-like current profile. The maximum Josephson currents ($I_{C0}$) were 2.118 mA for (a) and 1.118 mA for (b).

The sides of the junctions and the distances between their centers are the parameters $w$ and $l$, respectively, adequately indexed by the subscripts. The indexing of these parameters that we show in Fig. 1(a) is the one we use to calculate the terms of the equations that will follow few lines below for arrays with an even number of junctions: in this case, the $l$ that passes through the geometrical center of the array will be $l_1$ and the $w$ that follows $l_1$ on the right will be $w_1$. For arrays with an odd number of junctions, the situation is the opposite: one junction will occupy the geometrical center of the array and the width of this junction shall be $w_1$, while $l_1$ is the distance from the center of the junction to the center of its neighbour on the right. For such a system, developing concepts and equations taken from Refs. 1, 13, and 10, we have obtained a dependence of the maximum current upon the external magnetic field which puts no restrictions on the number of junctions of the arrays (even or odd) and on the distance and size of the junctions of the arrays. The modulation of the total Josephson currents through the arrays as a function of the applied external field $B$, in the case of uniform current density, for which the imaginary component of the spatial Fourier transform is zero,[10,13] reads

$$I_C(B) = \frac{I_{C0}}{\sum_{n=1}^{N} w_n^2} \left| (1-q)\Omega_1(B) + \sum_{j=2-q}^{m+1-q} \left[ (\Gamma_j(B) + \Theta_j(B)) \right] \right|, \quad (1)$$

where $I_{C0}$ is the maximum Josepson current in zero field of the whole array, $N$ is the number of junctions, and $m = [N/2]$, namely, $m$ is the integer part of $N/2$, while $q$ is also an integer and we set $q = 1$ for $N$ even and $q = 0$ for $N$ odd. The function $\Omega_1(B) = w_1^2 \left( \frac{\sin\left(\frac{\pi B d w_1}{\Phi_0}\right)}{\left(\frac{\pi B d w_1}{\Phi_0}\right)} \right)$, while $\Gamma_j(B)$ and $\Theta_j(B)$ are

$$\Gamma_j(B) = w_j^2 \frac{\sin\left(\frac{\pi B d w_j}{\Phi_0}\right)}{\left(\frac{\pi B d w_j}{\Phi_0}\right)}$$

$$\times \cos\left[\frac{\pi BD}{\Phi_0}\left(ql_1 + \sum_{k=1+q}^{j-1+q} l_k + l_{k+m-q}\right)\right], \quad (2)$$

$$\Theta_j(B) = w_{m+j}^2 \frac{\sin\left(\frac{\pi B d w_{m+j}}{\Phi_0}\right)}{\left(\frac{\pi B d w_{m+j}}{\Phi_0}\right)}$$

$$\times \cos\left[\frac{\pi BD}{\Phi_0}\left(ql_1 + \sum_{k=m+1}^{m+j-1} l_k + l_{k-m+q}\right)\right]. \quad (3)$$

In a forthcoming publication,[14] we will show in detail that Eqs. (1)–(3) are consistent extension of models present in the literature.[7,9,10,13] The comparison with the experimental results will herein tell us if the hypothesis of uniform current density is correct. It is easy to show that Eqs. (1)–(3) provide, for $N = 1$, the Fraunhofer pattern of a single junction [accounted for by the term $\Omega_1(B)$ indeed] and, for $N = 2$ and identical junctions, the two junction interferometer modulations.[1]

In the approach that we consider, the loops between the junctions are viewed as pieces of a junction that carry no current. Speaking in terms of a well known interferometer parameter, we work in the limit $\beta_L = 2\pi L I_0/\Phi_0 < 1$; here, $L$ is the inductance of the loop connecting the junctions (assumed to have the same critical current $I_0$), and $\Phi_0 = 2.07 \times 10^{-15}$ Wb is the flux quantum. When the two junctions in the loop have different Josephson currents, then we assume that the highest value of $\beta_L$ of the two connected junctions[1] must be less than unity. We will show later that the equation we propose can account even for the cases in which solutions had been found through direct solutions of the sine-Gordon model.[16]

In order to verify the predictions of Eq. (1)–(3), we designed a set of arrays which were then fabricated at HYPRES Inc, following a Nb-NAlOx-Nb—based trilayer protocol.[17,18] The samples had a critical current density $j_c = 140$ A/cm$^2$ and a Josephson penetration depth $\lambda_j = \sqrt{\frac{\Phi_0}{2\pi\mu_0 d j_c}} = 32\,\mu$m. All the measurements herein presented were performed at 4.2 K on samples electromagnetically shielded from the environment in order to provide stable and reproducible results. The external magnetic field was provided by a solenoid. All the tested samples had excellent current-voltage characteristics which were recorded in our data acquisition system as a function of the applied magnetic field and from these the diffraction patterns were obtained.

In Fig. 2(a), we show the results for a 15 junction array: in this case, speaking in terms of Eq. (1), $m = [N/2] = 7$. The areas of the junctions are nominally identical (10 $\mu$m × 10 $\mu$m), with a spread in the fabrication parameters that we estimate less than 1%. The junctions are all equally spaced 5 $\mu$m and therefore all $l_i = 15\,\mu$m. Here, we see that the generated interferometric oscillations have 15 as a repetition period and the full line through the data, representing the prediction of Eq. (1), provides a good fitting. From Fig. 2, we have an interesting conclusion in terms of device-oriented applications:[12] we see that it is possible to distribute one fluxon over the whole length $W$ of the array, which happens for $B_W = 0.4$ G in Fig. 2(a), while the flux connecting neighbour junctions remains very small due to the fact that connecting inductances are small as well. Due to the low inductance of the loops connecting the junctions, the spatial distribution of the flux over the array in low fields is not far from that described in Sec. 5.2 of Ref. 1 for continuous junctions. Instead for $B_l = 5.75$ G, we have one flux-quantum for every loop connecting the junctions. For $B_J = 11.4$ G (not shown in the figure), the envelope of the maxima goes to zero and we will have one flux quantum per junction. Experimental uncertainties have values within the dots in the figures.

All our patterns were very symmetrical for inversion of the magnetic field direction and an example is given in Fig. 2(b) for a parallel array of 8 identical (10 $\mu$m × 10 $\mu$m) junctions all spaced 30 $\mu$m, and therefore with all $l_i = 40\,\mu$m [see the microscopy image in Fig. 2(c)]. The lines through the experimental dots represent the predictions of Eq. (1) with m = [N/2] = 4, all $w_i$ equal were to 10 $\mu$m, and all $l_i$ were equal to 40 $\mu$m. The parallel connection of 8 junctions generates now a repetition period equal to 8 oscillations which is a typical feature of Eq. (1). From the fitting equation, we extrapolate the zero of the modulations corresponding to have one flux-quantum in each single junction which occurs for $B_J = 11.4$ G [the same of the 15 junction array of of Fig. 2(a) since the width of the junctions is the same]. From the equation $\Phi_0 = B_J w d$ and $d = 2\lambda_{Nb} + t_{ox}$, we calculate $d = 180$ nm which is consistent with the values given by HYPRES (London penetration depth $\lambda_{Nb} = 90$ nm). The thickness $t_{ox}$ of the junction oxide ($Al_2O_3$), few nanometers, is neglected.

Each "small" modulation in Fig. 2 corresponds to have flux quanta $\Phi_0$ sequentially penetrating through the whole area available for penetration over the length $W$. The distance between the high maxima, occurring after eight modulations, is determined by the field $B_l$ responsible for trapping one flux-quantum in every loop. For the field $B_l = 1.84$ G, corresponding to the distance between the two wide amplitude maxima, we extract from the condition $B_l l D = \Phi_0$, taking $l = 40\,\mu$m, we get $D = 280$ nm which is fully consistent with the fabrication process, assuming for $D = 2\lambda_L + t_{SiO2}$ the sum of twice the London penetration depth of niobium (180 nm) and 100 nm of $SiO_2$ (see Fig. 1). From the above estimates of $D$, we have calculated $\beta_L = 0.16$ and $\beta_L = 0.5$ for the 15 junction array and the 8 junction array, respectively.

In Fig. 3, we validate our model in the case of disuniformity between junction areas and distances. This specific case has also been worked out through a direct model of the sine-Gordon equation with appropriate boundary conditions.[16] In the top part of Fig. 3(a), we show a sketch of a top view of of the designed array for which we followed indeed the design parameters suggested in Ref. 16, while the photo at the bottom shows a fabricated sample. The curve fitting the data in Fig. 3(b) has been obtained from Eq. (1) by setting adequately the parameters corresponding to the sample sketched in Fig. 3(a). In the "central" part of the modulations, our result is essentially undistinguishabe from the predictions of Ref. 16 (see Fig. 7 of that paper) where, however, the diffraction effects were not considered and the side maxima had the same height of the central one. Taking into account the diffraction effects through Eq. (1), we can fit the secondary maxima as shown in Fig. 3(c). The result is that, even for such a peculiar structure, we can provide a good account of the static behaviour. An interesting counterpart of this geometry is that the interferometric modulations due to the five junctions have disappeared. We also note that all along the diffraction pattern of Fig. 3(b), resonances were not observed in the current-voltage characteristic of the array; this feature (observed even on other arrays having juctions having slightly different sizes) could offer advantages in terms of stability, and noise reduction, of magnetic field detectors and digital devices.

In Fig. 4(a), we show the modulations of a 20 junction (all equally spaced 2 $\mu$m, $\beta_L = 0.121$) interferometer, fitted by our Eqs. (1)–(3) while in the inset, we show all the Fiske resonances traced sweeping slightly the external magnetic field around 1.9 G. As in the other cases, we see that Eq. (1) provides an excellent fit to the data confirming the hypothesis of uniform current density. The voltage spacing between the resonances is 40 $\mu$V: relating this value to frequency $f$ by the Josephson ac relation, given the resonance condition $f = \bar{c}/2W$ ($W = 240\,\mu$m), a speed of light along the one-dimensional structure $\bar{c} = 0.03c$ is calculated. This result agrees with previous data[15,19] confirming that resonances develop over the whole length $W$ and that makes even dynamical sense to consider the arrays like a single junction.

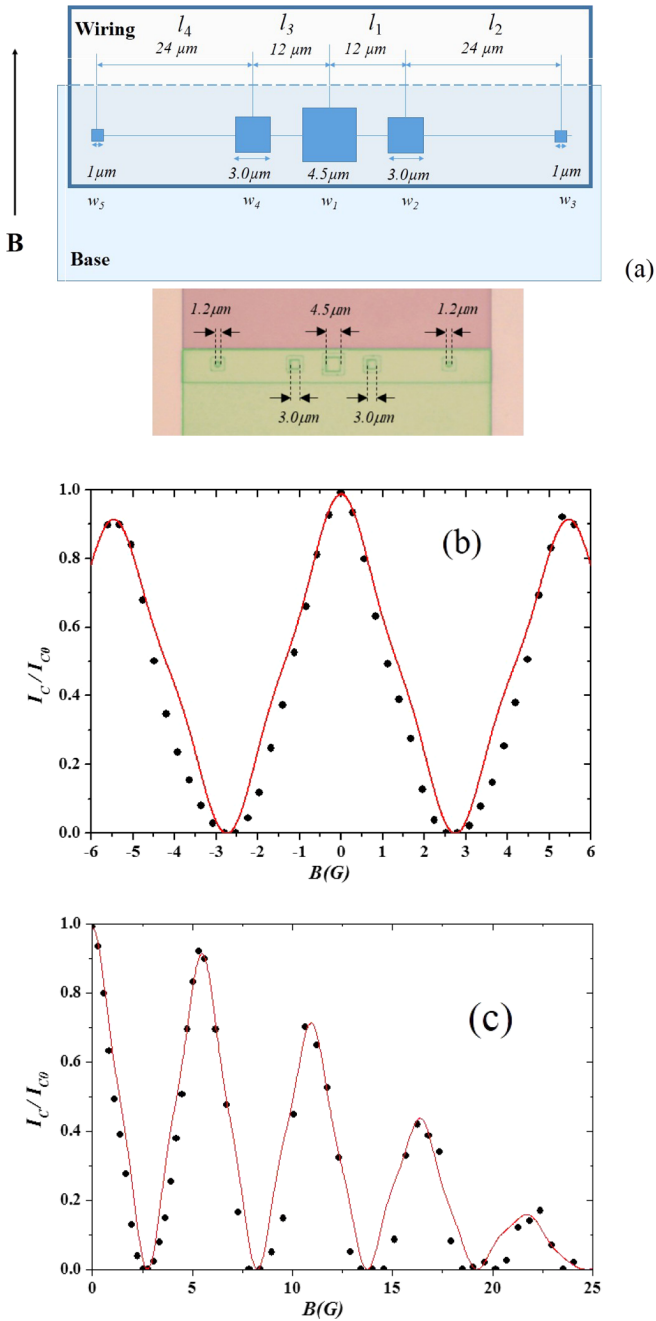

FIG. 3. (a) Top: (sketch) of an array with different areas of junctions and different spacings. Bottom: optical microscope photo of the fabricated device. Note that the smallest junctions have a circular shape with a 1.2 $\mu$m diameter; (b) fitting of the experimental data (circles) by Eq. (1); and (c) the diffractive effect taken for increasing values of the field. Indexing of the junctions is what we use for arrays with an odd number of junctions. The maximum Josephson current was 49 $\mu$A.

The singularities shown in Fig. 4 extend up to 700 $\mu$V with a very regular spacing indicating the stability of the resonant excitations. From the Josephson *ac* relation, the spanned voltage range of Fig. 4 corresponds roughly to a frequency interval between 150 and 340 GHz; moreover, we estimate that the available power on the resonances is in the range of several hundreds of nanowatts. For devices operating with specific characteristics in the terahertz gap, active components with regular properties like those shown in Figs. 2 and 4 open encouraging perspectives.

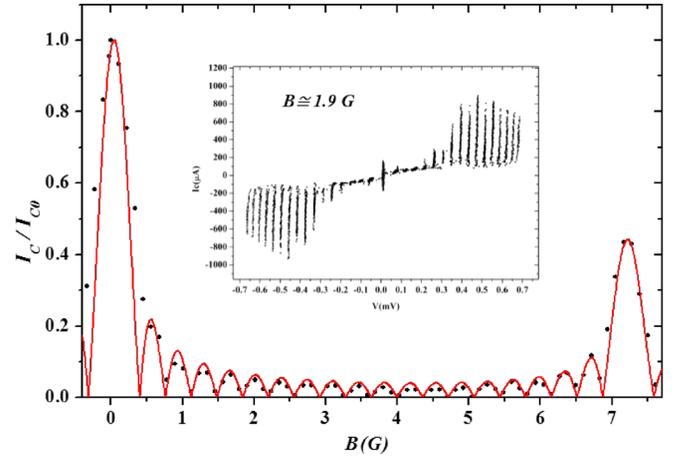

FIG. 4. (a) The modulations of a regular array of 20 junctions interpolated by our model Eq. (1); the inset shows the Fiske modes of the 20 junction array traced sweeping slightly the magnetic field around $B \cong 1.9$ G. The maximum critical current of the array was 2.831 mA.

In conclusion, we have explained the current modulations of one dimensional inhomogeneous arrays of Josephson junctions in terms of a model describing arrays as single junctions with a position-dependent current. Our model equations allow predicting the response of arrays engineered for specific goals, a relevant feature in the metadevices perspective. Indeed, nowadays systems integrating very large arrays of Josephson of low $\beta_L$ interferometers have been engineered for rf applications.[20] For these systems, numerical simulations are somewhat prohibitive and our analytical approach could provide hints for understanding the response and improving the device performance.

The financial support of the *INFN* (Italy) through the project *FEEL* is acknowledged.